\documentclass[nobibnotes,nofootinbib,twocolumn]{revtex4-1}
\usepackage[english]{babel}
\usepackage{ucs}
\usepackage[utf8x]{inputenc}
\usepackage{amsmath,amssymb,bm,cancel}
\usepackage{graphicx}
\usepackage{subcaption}
\usepackage{color}
\graphicspath{{figures/}}
\def \grad {\boldsymbol{\nabla}}
\def \div {\grad\cdot}
\def \curl {\grad\times}

\def \gammalr {\gamma_{\text{LR}}}
\def \gammaext {\gamma_{\text{ext}}}
\def \gammavde {\gamma_{\text{VDE}}}
\def \fext {\bar{F}_{\text{ext}}}
\def \aext {\bar{A}_{\text{ext}}}
\def \litwo {\text{Li}_2}

\newcommand{\amendment}[1]{#1}
\begin{document}
\title{Algebraic motion of vertically displacing plasmas}
\author{D Pfefferlé$^1$}
\author{A Bhattacharjee$^1$}
\affiliation{$^1$Princeton Plasma Physics Laboratory (PPPL), Princeton 08540 NJ, USA}
\begin{abstract}
  The vertical motion of a tokamak plasma is analytically modelled
  during its non-linear phase by a free-moving current-carrying rod
  inductively coupled to a set of fixed conducting wires or a
  cylindrical conducting shell. The solutions capture the leading term
  in a Taylor expansion of the Green's function for the interaction
  between the plasma column and the surrounding vacuum vessel. The
  plasma shape and profiles are assumed not to vary during the
  vertical drifting phase such that the plasma column behaves as a
  rigid body. In the limit of perfectly conducting structures, the
  plasma is prevented to come in contact with the wall due to steep
  effective potential barriers created by the induced Eddy
  currents. Resistivity in the wall allows the equilibrium point to
  drift towards the vessel on the slow timescale of flux
  penetration. The initial exponential motion of the plasma,
  understood as a resistive vertical instability, is succeeded by a
  non-linear ``sinking'' behaviour shown to be algebraic and
  decelerating. The acceleration of the plasma column often observed
  in experiments is thus concluded to originate from an early sharing
  of toroidal current between the core, the halo plasma and the wall
  or from the thermal quench dynamics precipitating loss of plasma
  current.
\end{abstract}
\maketitle
 \section{Introduction}
 \label{sec:intro}
 To reach higher confinement and transport performances, tokamak
 plasmas are given an elongated shape by the application of external
 magnetic fields from two major current carrying coils (divertor
 coils) located at the top and bottom of the device. This setup is
 inherently unstable to vertical displacements so that, without the
 presence of conducting wall structures and active feedback control,
 the plasma rapidly moves up or down the vacuum chamber before coming
 in contact with the first wall~\cite{lazarus-1990}. The rapid
 transfer of current that occurs mostly at the end of a Vertical
 Displacement Event (VDE) leads to immense stress on the conducting
 structures and can cause severe damage to the reactor's vacuum
 vessel~\cite{pick-1991,riccardo-2003,neyatani,gerhardt-2012}. Prevention
 and mitigation of such disruptive events is a central issue for ITER
 operation~\cite{hender,lehnen-2015}. \amendment{Since timing is
   critical, every bit of information that is processed early into the
   VDE can be useful to determine the safe shutdown of a disrupting
   discharge.}

A basic way to understand the characteristic motion of a plasma column
during a VDE is via a simple magneto-dynamical model
\citep{freidberg}, involving a set of rigid conducting wires and thin
shells to represent the coupling between the plasma and the
surrounding conducting structures. \amendment{Several studies of the vertical
instabilty in tokamak plasmas rely on such \emph{wire-model} to obtain
linearised circuit and motion
equations~\citep{jardin-1982,jardin-1986,sayer-1993}. These studies
provide valuable insight into the challenge of tokamak control and
feedback system design. Natural extensions of this method have since
been developed where the plasma response is described by linear MHD
equations inductively coupled to wire or finite element
representations of the vacuum
vessel~\cite{marsf,starwall,pet,villone-2015}. The wire-model is also
the basis for several axisymmetric tokamak
codes~\cite{jardin-1986,dina,maxfea,pet}, which simulate the nonlinear
plasma behaviour as a sequence of Grad-Shafranov equilibria. In those
codes, the time-evolution of profiles and reference flux-surfaces is
prescribed, since the plasma dynamics is not resolved. The numerical
results obtained with those tools yield realistic estimates of wall
currents and forces, but are valid only within the predetermined
(ad-hoc) scenario. The self-consistent modelling of transient tokamak
phenomena is becoming tractable thanks to the increasing power of
high-performance computing and the deployment of large 3D resistive
MHD simulations \cite{m3dc1,nimrod,jorek}. A serious conceptual
limitation behind those codes is the way in which the region beyond
the last-closed flux-surface is treated as a cold and low density
plasma, which can falsly dominate the VDE evolution by acting as a
conducting line-tied shell \cite{pfefferle-2018}. In contrast,
wire-models treat the region beyond the last-closed flux-surface as
true vacuum, which is more consistent and realistic in the early
stages of a VDE. The assessment of MHD activity within the framework
of wire-models requires a suitable parametrisation of internal fluid
displacements and seems possible. Including plasma-wall contact or
three-dimensional effects is also under consideration.}

\amendment{While plenty of linear analytic solutions for the vertical instability
exist in the litterature, a \emph{non-linear} analytic description of
the drifting phase of a VDE seems not to exist. In most
interpretations, the $n=0$ vertical mode is considered either stable
(oscillatory) or unstable (exponentially growing). The set of linear
ordinary differential equations (ODEs), as in section
\ref{sec:linear_analysis}, characterises the behaviour of the plasma
column at the onset of a VDE, but provides an incomplete picture of
the non-linear electro-magneto-mechanical physics taking place on a
longer timescale, especially in the intermediate drifting phase before
the plasma comes in contact with the wall. It is indeed shown in this
paper that, if there is no sharing of current between the plasma and
the wall, the motion of the plasma column quickly ceases to be
exponential as it approaches the conducting wall and will always
exhibit a phase of algebraic deceleration. The slow speed of the VDE
with respect to the Alfvén frequency suggests that the $n=0$ resistive
vertical instability evolves non-linearly as a relaxation of the
system's equilibrium position. This interpretation of the evolution of
instabilities that arise during a disruption in the vicinity of a
stabilising conducting wall remains largely valid for non-axisymmetric
modes~\cite{zakharov-2008,gerasimov-2015}. The conducting structure
surrounding the plasma acts in effect as a high-pass filter, damping
out fast Alfvénic activity. With our reduced model and its direct
numerical extension to any wall geometry, one is able to predict the
time-evolution of the current centroid position as well as toroidal
wall currents via one single non-linear ordinary differential
equation. Real-time comparison with experimental traces could be used
to identify the inflexion point of the VDE and the moment when the
plasma column comes in contact with the wall. Ongoing numerical
modelling efforts may also benefit from the existence of simple
analytic formulae for benchmarking and code verification purposes.}

The paper is organised as follows. Section \ref{sec:single_wire}
illustrates some basic ideas and observations from a toy model where
the wall and plasma are each represented by a single wire. The outcome
of a typical linear analysis is discussed in section
\ref{sec:linear_analysis} and an analytic solution to the slow
non-linear motion of the plasma column is proposed in section
\ref{sec:resdecay_single}. Section \ref{sec:plasma_wire} justifies
under which circumstances the plasma can be approximated by a rigid
rod. In section \ref{sec:multiple_wires}, a convenient framework to
describe the wall as a collection of wires using a variational
principle is derived. A general ordinary differential equation for the
algebraic VDE motion is deduced in section \ref{sec:multiple_decay}
for regimes relevant to present-day tokamaks, in which the wall
current decay time is much slower than Alfvénic but faster than the
Ohmic dissipation of plasma current. This ODE is solved analytically
for the case where the wall is represented by two wires on each side
of the plasma. In section \ref{sec:cylindrical_shell}, the evolution
equations for surface currents on a cylindrical resistive shell are
obtained. Leaning on the general ODE of \ref{sec:multiple_decay}, the
algebraic motion of the plasma column is given an analytic solution in
the presence of this surrounding shell in section
\ref{sec:resdecay_shell}. Several applications of our findings and
possible extensions are discussed in section \ref{sec:conclusion}.
\section{Basic idea and observations from a single wire wall model}
\label{sec:single_wire}
\amendment{The simplest description of the vertical motion of a plasma
  column during a VDE is given by the following elementary model
  \citep{freidberg,jardin-1982}}. Consider two straight parallel wires
of long length $l$. The plasma-wire moves freely along the vertical
$z$-axis, initially positioned at $z(0)=0$, and is traversed by a
constant current $I_p$ flowing in the (toroidal) $y$-direction. The
second wire, representing the conducting wall, is fixed at $z=a>0$ and
carries a time-varying current of $I_w(t)$.

The plasma-wire generates at the location of the wall-wire a poloidal
magnetic field $\bm{B}_p(t) = \hat{\bm{x}} \mu_0
I_p/2\pi(a-z(t))$. The wall-wire produces
$\bm{B}_w(t) = -\hat{\bm{x}}\mu_0 I_w(t)/2\pi(a-z) $ at the location
of the plasma-wire, creating an attractive force when the currents
have the same sign (Ampère's force). The equation of motion for the
plasma-wire is written
\begin{equation}
  \label{eq:eom_z}
m\frac{d^2z}{dt^2} = \frac{\mu_0 l I_p}{2\pi} \frac{I_w(t)}{a-z} + F_{ext}(z)
\end{equation}
where $m$ is its mass and $F_{ext}(z) = -\grad V_{ext}$ is a
time-independent external force, from the control coils for example.

The wall wire being an inductor-resistor, its current is generated by
the electromotive force
\begin{equation}
  \label{eq:eom_I}
  L_w \frac{dI_w}{dt} + R_w I_w = \int (\bm{v}_{\text{rel}}\times \bm{B}_p)\cdot d\bm{l}
  = -\frac{\mu_0 lI_p}{2\pi(a- z)}\frac{dz}{dt}
\end{equation}
where $\bm{v}_{rel} = \bm{v}_w-\bm{v}_p$, $R_w$ is the wall resistance
and $L_w$ its self-inductance.

It is convenient to adimensionalise our variables by introducing the
normalisation
\begin{align}
\bar{z} &=z/a &
 \bar{I}_w &= I_w/I_p   &
 L_c &= \frac{\mu_0 l }{2\pi}  &
 \tau &= t/t_c.
\end{align}
The characteristic (short) timescale is chosen to be Alfvénic
\begin{equation}
t_c = \sqrt{ \frac{ m a^2}{L_c I_p^2}}
 = \frac{\sqrt{\mu_0 \rho_c}}{B_c}a  =  \frac{a}{v_A}
\end{equation}
where $B_c = \mu_0 I_p/2\pi a$ is the initial poloidal field generated
by the plasma column at the wall, $\rho_c = m/l2\pi a^2$ half the
plasma mass density if it were filling the vacuum chamber uniformly
and $v_A=B_c/\sqrt{\mu_0 \rho_c}$ the associated Alfvén velocity. For
typical plasmas\footnote{$I_p=1 MA$, $n=10^{20} m^{-3}$, $a=1m$.}, the
characteristic timescale is thus of the order of $t_c \sim 10^{-6}
s$. The wall inductance and resistance are normalised by
\begin{align}
\bar{L}_w &=   L_w /L_c   &
\bar{R}_w & = R_w t_c/L_c
\end{align}
where the normalised conductance $\bar{R}_w^{-1}= S_w (A/2\pi a^2)$ is
equal to the wall Lundquist number $S_w = \mu_0 a v_A/\eta_w$ times
its cross-section $A$ divided by twice the area of the vacuum chamber
$\pi a^2$. A typical iron wall Lundquist number is $S_w\sim 10^7$,
which makes $\bar{R}_w \sim 10^{-5}-10^{-3}$ depending on the
cross-section. The wall's self-inductance is expected to reach
$\bar{L}_w\sim 1-10$, depending on the shape of the structure
considered.

The following four normalised frequencies/growth rates will be
identified
\begin{align}
  \label{eq:timescales}
  \Omega & = \frac{1}{\sqrt{\bar{L}_w}} \sim 1 &
 \gammalr &= \frac{\bar{R}_w}{\bar{L}_w} \ll 1\\
 \gammaext & 
             < \Omega &
 \gammavde &= \bar{R}_w \gammaext^2 \ll 1.
\end{align}
Their significance will become evident shortly, but $\Omega$ is the
oscillation frequency around the equilibrium point, $\gammaext$ is the
growth rate induced by the unstable external potential, $\gammalr$ is
the ``L over R'' time or decay rate of wall currents, $\gammavde$ is
the actual growth rate of the VDE. The ratio between the driving term
and oscillation frequencies, which also happens to be the ratio
between the wall time and VDE time, plays an important role in our
analysis. For reasons detailed in the results, this ratio is smaller
than unity (if not much smaller)
\begin{equation}
  s^2 
 = \frac{\gammaext^2}{\Omega^2} = \frac{\gammavde}{\gammalr} < 1,
\end{equation}
and suggests the following ordering of available timescales
\begin{equation}
  \label{eq:ordering}
  \gammavde < \gammalr \ll  \Omega \sim 1.
\end{equation}

The adimensional equation of motion and circuit equation are written
as a system of three coupled first-order ODEs
\begin{equation}
  \begin{pmatrix}
    \dot{\bar{z}} \\
    \dot{v} \\
    \dot{\bar{I}}_w
  \end{pmatrix}
   =
  \begin{pmatrix}
    v \\
    \frac{\bar{I}_w}{1-\bar{z}} + \fext(\bar{z})\\
     -\frac{\Omega^2 v}{1-\bar{z}} - \gammalr \bar{I}_w
     \end{pmatrix}
  = \bm{F}(\bar{z},v,\bar{I}_w)
  \label{eq:eoms}
\end{equation}
where $\dot{X} = dX/d\tau$ and $\fext(\bar{z})=t_c^2
F_{\text{ext}}(z)/m a \sim \gammaext^2\bar{z}$ is the normalised external driving force.
\subsection{Stable points and linear analysis}
\label{sec:linear_analysis}
The wall resistivity makes the system dissipative (sink term), as
concluded from the non-zero trace of the Jacobian matrix,
$\text{tr}[D\bm{F}] = \partial\dot{X}^i/\partial X^i =
-\bar{R}_w/\bar{L}_w = -\gammalr$ and from the assumption that the
external force conservative.

In the absence of external fields ($\fext\equiv 0$), an infinite
amount of extremal points are found on the $(\bar{z}_e,0,0)$ line for
which the force term $\bm{F}(\bar{z}_e,0,0)=0$ vanishes. Their stability
is assessed by finding the eigenvalues of the Jacobian matrix
$D\bm{F}(\bar{z}_e,0,0)$, i.e.
\begin{align}
  \gamma_0 &= 0   \label{eq:eigenzero}\\
  \gamma_\pm &= \pm i \sqrt{\frac{\Omega^2}{(1-\bar{z}_e)^2}-\frac{\gammalr^2}{4}} -\frac{\gammalr}{2}\label{eq:eigenvalues}.
\end{align}
Without external forcing, the system is concluded to be globally
stable with a real negative growth rate (damping) inversely
proportional to the wall time, highlighting the stabilising role of
the wall. Neglecting resistivity, the frequency of small vertical
oscillations is Alfvénic (of order unity compared to characteristic
time $t_c$) and increases as the plasma is brought closer to the wall
wire.

In the presence of the external force from the divertor coils,
expressed in the form of equation (\ref{eq:divertor_field}), the only
equilibrium point is the origin of
$(\bar{z},v,\bar{I}_w)=(0,0,0)$. Its stability is assessed via the
cubic eigenvalue equation
\begin{equation}
    \gamma^3 + \gammalr \gamma^2 + (1-s^2)\Omega^2\gamma - \gammavde \Omega^2=0
\end{equation}
In the limit of highly conducting wall structures, the growth rates of
equation (\ref{eq:eigenzero}-\ref{eq:eigenvalues}) are replaced by
\begin{align}
  \gamma_0   &= \frac{\gammavde}{1- s^2} +  O(\gammalr^2)  \label{eq:drive_eigenzero}\\
  \gamma_\pm &=  \pm i \Omega\sqrt{1-s^2} -\frac{\gammalr}{2(1-s^2)}+ O(\gammalr^2). \label{eq:drive_eigenvalues}
\end{align}
The frequency (damping) of oscillatory modes $\gamma_\pm$ is weakened
(strengthened) through the factor $s^2$. A slow positively growing
mode, $\gamma_0$, emerges due to the forcing, generating the initial
exponential phase of the VDE.
\subsection{Non-linear dynamics in the perfectly conducting wall limit}
Neglecting the system's dissipation, i.e. considering the limit
$\gammalr\rightarrow 0$, equations (\ref{eq:eoms}) are integrable. One
obtains the following quadrature 
\begin{equation}
  \bar{I}_w-\bar{I}_{w,0} = \Omega^2\ln \left|\frac{1-\bar{z}}{1-\bar{z}_0}\right|
\end{equation}
and
\begin{equation}
\frac{1}{2}v^2 + \underbrace{\frac{\Omega^2}{2} \left( \ln\left|\frac{1-\bar{z}}{1-\bar{z}_0}\right| +\frac{\bar{I}_{w,0}}{\Omega^2}\right)^2+ \bar{V}_{\text{ext}}(\bar{z})}_{V_{\text{eff}}(\bar{z})} 
= \frac{1}{2}v_0^2 + \frac{\bar{I}_{w,0}^2}{2\Omega^2}  \label{eq:curve_zv}
\end{equation}
where $\bar{V}_{\text{ext}}' = -\fext \propto -\gammaext^2\bar{z}^2/2$ defines the adimensional external
potential. The first term of the effective potential $\ln| 1-\bar{z}|^2$
is seen to have a global minimum.

\amendment{Adding a polynomial function to it might dominate at large $\bar{z}<0$
but certainly does not remove the logarithmic divergence when the
plasma approaches the wall at $\bar{z} \rightarrow 1$. The wire model
suggests that it is impossible to compress the plasma against the wall
on Alfvénic timescales, a conclusion that is well known from initial
studies of tokamak equilibria~\cite{mukhovatov-shafranov}. The only
way the two wires can come in contact is through a
resistive/dissipative process in conjunction with an external driving
potential. An equilibrium position exists on Alfvénic timescales no
matter the initial conditions, unless $\gammaext \geq \Omega$. In the
latter case the divertor field annihilates the minimum of the
effective potential (\ref{eq:curve_zv}) and the plasma wire is forced
to move away from the wall. Since this case is not particularly
interesting, we will assume that $s^2=\gammaext^2/\Omega^2 <1$. The
effect of resistivity is listed i) to damp the fast oscillations on
``L over R'' timescales, so that the system reaches the position of
minimum effective potential, ii) to gradually shift the equilibrium
position over the VDE timescale in the presence of a driving field.
}
\subsection{Resistive decay of wall currents and characteristic
  vertical plasma motion}
\label{sec:resdecay_single}
\amendment{When external fields are applied and the induced wall currents have
time to dissipate, the system's equilibrium position changes as the
minimum of the effective potential is brought closer to the wall. The
plasma is thus pictured to ``sink'' across the external potential, in
much the same way as a cushion filled with air would slowly deflate
due to a tiny puncture. The next paragraph describes how fast the
vertical motion is throughout this non-linear relaxation process.}

Given the ordering of equation (\ref{eq:ordering}), the evolution of
the equilibrium position, $\bar{z}_e(\tau)$, of the basic two-wire
system is conveniently found by requiring that the induced wall
currents exactly compensate the external force such that the plasma is
constantly in force balance
\begin{align}
  \label{eq:resistive_regime}
    \dot{v}& = \frac{\bar{I}_w}{1-\bar{z}} + \fext = 0 &
\iff
 \bar{I}_w& =-(1- \bar{z}) \fext 
\end{align}
%
%
where the subscript $\bar{z}_e\equiv \bar{z}$ was dropped for
convenience. \amendment{This assumption essentially justifies the
  calculation of a sequence of Grad-Shafranov equilibria in tokamak
  evolution codes such as TSC~\cite{jardin-1986} and
  DINA~\cite{dina}.} Substituting (\ref{eq:resistive_regime}) in
equation (\ref{eq:eoms}) for the wall current, the following
differential equation is obtained for the equilibrium position
\begin{align}
  \label{eq:resistive_equation}
\dot{\bar{z}}\left\{\frac{1}{(1-\bar{z})^2\fext/\gammaext^2} - s^2\left[\frac{\fext'}{\fext} -\frac{1}{1-\bar{z}}\right]\right\}=  \gammavde
\end{align}
This equation assumes that the fast Alfvén dynamics has relaxed and
only the slow drifting motion, $\bar{z}_e(\tau)$, of the equilibrium
position is traced. It is understood that the plasma acceleration
(change in velocity) is of order
$\dot{v} \sim O\left(\gammavde^2\right)$ in this regime. The timescale
separation implied in equation (\ref{eq:resistive_equation}) is a
justified limit to understand the non-linear ``sinking'' of the plasma
due to current dissipation in the wall as it ``feels'' the external
potential without having to numerically solve the full (Alfvénically
coupled) system of equations.

\paragraph{External field from divertor coils:} In the presence of the
external force produced by the divertor coils
(\ref{eq:divertor_field}), the differential equation
(\ref{eq:resistive_equation}) can be cast as
\begin{equation}
  \dot{F}(\bar{z}) - h^{-2}\dot{H}(\bar{z}) - s^2 \dot{K}(\bar{z}) = \gammavde
\end{equation}
for which an implicit solution $\bar{z}(\tau,h)$ is readily obtained 
\begin{equation}
  \label{eq:divertor_single}
F(\bar{z}) - h^{-2} H(\bar{z}) - s^2 K(\bar{z}) = (\tau-\tau_0)/\tau_{VDE}
\end{equation}
where $\tau_{VDE} = \gammavde^{-1}$ and
\begin{align}
  F(\bar{z})& =   \frac{\bar{z}}{1-\bar{z}} +  \ln\left|\frac{\bar{z}}{1-\bar{z}}\right| \\
  H(\bar{z})&=  \frac{\bar{z}}{1-\bar{z}} + \ln|1-\bar{z}| \\
  K(\bar{z})&= \ln\left|\frac{\bar{z}(1-\bar{z})}{1-h^{-2} \bar{z}^2}\right|.
\end{align}
The inverted graph of the implicit function (\ref{eq:divertor_single})
is shown on figure \ref{fig:unstable_single}) for several values of
$s^2$ and $h$. The equilibrium position is concluded to evolve along
the reference time $T=(\tau-\tau_0)/\tau_{\text{VDE}}$, which depends
weakly on the wall inductance. The shape of the solution is affected
by the frequency ratio between the driving force and the Eddy
potential, $s^2=\gamma^2_{ext}/\Omega^2=\gammavde/\gammalr$ as well as
by the position of the divertor coils parametrised by $h=z_c/a$. Close
to the unstable position, the behaviour $\forall h > 1$ is exponential
with the initial growth rate equal to the leading order linear
$\gamma_0$ of equation (\ref{eq:drive_eigenzero}), as highlighted by
the dashed lines on figure (\ref{fig:unstable_single}),
\begin{align}
  \label{eq:init_growth}
  \frac{\dot{\bar{z}}}{\bar{z}} &\overset{\bar{z}\rightarrow 0}{\longrightarrow} \frac{\gammavde}{1-s^2} &\Rightarrow&&
 \bar{z}(T)& \overset{T\rightarrow -\infty}{\longrightarrow} \bar{z}_0 e^{T/(1-s^2)}
\end{align}
Near the wall, the motion becomes a slow $1/T$ decay and the
instantaneous growth rate monotonically decreases to reach zero. This
algebraic deceleration resulting from the non-linear relaxation
process is an important outcome in all models of this paper.

The point of inflexion, which is an experimentally relevant measure of
the start of the decelerating phase, is found by solving
$\ddot{\bar{z}}(\bar{z}_*)=0$. On figure \ref{fig:unstable_single}),
it is illustrated for each value of $s^2$ and $h$ by a circle. The
largest value the inflexion point reaches is
$\bar{z}_*(h\rightarrow 1,s^2\rightarrow 0)=\sqrt{2}-1\cong 41.4\%$,
which is a purely geometric result. \amendment{Counter-intuitively, inflexion of
the vertical motion occurs earlier when the external drive $s^2$ is
increased. In other words, the initial exponential phase extends over
a shorter distance when $\gammavde$ becomes comparable to
$\gammalr$. In experiments and simulations however, the exponential
phase seems to be span over the entire VDE, suggesting that the
characteristic VDE time is much longer than the wall time. Whether
this regime is representative of future devices like ITER, where the
wall time will be remarkably long due to the thickness of the
conducting vacuum vessel, remains to be determined.}

The potential created by the divertor coils tends to a parabola in the
limit where the distance $h\rightarrow \infty$. In this case, an
explicit solution in terms of the Lambert W function is found for
$s^2\rightarrow 0$,
\begin{equation}
  \bar{z}(T; h\rightarrow \infty, s^2\rightarrow 0) = \frac{1}{1+1/W( e^T)}.
\end{equation}
This curve is depicted on figure \ref{fig:unstable_single}) by the
dashed black line.
\subsection{Validity of the wire-model}
\label{sec:plasma_wire}
\amendment{Due to its simplistic treatment of the plasma, the wire-model
presented in the previous section is limited in achieving a
comprehensive description of the vertical drift. The model
nevertheless comes a long way in explaining the scaling and
qualitative behaviour of the plasma column during a VDE. In this
section, we discuss its legitimacy and carefully detail the effects
neglected.}

If there is no contact between the plasma and the wall, the transfer
of current is purely inductive. The coupling between poloidal currents
being much weaker than between toroidal currents, we can safely focus
on poloidal magnetic fields and the integrated Lorentz force caused by
toroidal currents only. The total poloidal magnetic field is
decomposed into the plasma, the wall and the external coil components,
and expressed in terms of respective poloidal fluxes as
$\bm{B} = \bm{B}_p + \bm{B}_w + \bm{B}_{\text{ext}} = \grad (\Psi_p +
\Psi_w + \Psi_{\text{ext}})\times\grad\varphi$.

In the wall, toroidal currents are driven by the time-variation of the
poloidal flux. The contribution from the plasma can be evaluated,
knowing that
$\Delta^*\Psi_p = R^2\div(\grad\Psi_p/R^2) = R j_\varphi$, by the
Green's function method \citep{jardin-1986} as
\begin{align}
  \Psi_p(\bm{x},t) =\frac{\mu_0}{2\pi} \int_{S} G(\bm{x}',\bm{x}) j_\varphi(\bm{x}',t) d\bm{x}'
\end{align}
where $\bm{x} = (R,Z)$ is a point on a poloidal plane in the wall and
$\bm{x}'$ a point in the plasma, $d\bm{x}'=dR'dZ'$, $S$ a poloidal
surface area large enough to enclose the plasma at all times,
$j_\varphi = \bm{j}_p\cdot \hat{\bm{e}}_\varphi$ the toroidal plasma
current density and $G$ the Green's function satisfying
$\Delta^*G = 2\pi R\delta(R-R')\delta(Z-Z')$. The total plasma
current, the position of the current centroid and the quadrupole
tensor are defined respectively
\begin{align}
I_p(t) & = \int_{S} j_\varphi(\bm{x},t)d\bm{x}' \\
\bm{x}_p(t) &= \frac{1}{I_p}\int_S \bm{x} j_\varphi(\bm{x},t)d\bm{x}' \\
\bm{K}(t) &=  \frac{1}{I_p}\int_S (\bm{x}-\bm{x}_p)(\bm{x}-\bm{x}_p) j_\varphi(\bm{x},t)d\bm{x}'.
\end{align}
Taylor expanding the Green's function about the ``centre-of-current'',
$\bm{x}_p(t)$,
the plasma poloidal flux evaluated at the wall is approximately
\begin{align}
  \label{eq:psipatwall}
  \Psi_p(\bm{x},t) = \frac{\mu_0 I_p}{2\pi}\left[ G(\bm{x},\bm{x}_p) + \frac{1}{2}\bm{K}:\grad\grad G(\bm{x},\bm{x}_p) + \ldots\right].
\end{align}
The first term represents a current-carrying wire at the location of
the current centroid. In general, the position of the current
centroid, the centre-of-mass and the magnetic axis do not coincide,
but the difference is expected to be proportional to the inverse
aspect-ratio, the ellipticity and other shaping
parameters. Triangularity has been reported to affect the linear
growth rate in the context of positional
control~\cite{ambrosino-albanese}. However, during a VDE, the plasma
column rapidly loses its D-shape and appears to have a compact and
circular shape on fast cameras. For the purpose of our discussion,
those effects are assumed to be sub-dominant and the three positions
are considered to coalesce. The second term in (\ref{eq:psipatwall})
is a quadrupole correction, neglected in this work on the basis that
the shape of the plasma is sufficiently close to circular and the
plasma is sufficiently far away from the wall that it yields again a
sub-dominant contribution \cite{jardin-1986}.

The plasma column moves due to the Lorentz force of its current times
the magnetic fields caused by currents in the wall and external
coils. The time-variation of the total plasma momentum is
\begin{align}
  \frac{d \bm{P}}{dt}& = \frac{d}{dt}\int_{V_p(t)} \rho\bm{V} d\omega
  = \int_{V_p(t)} \rho\frac{d \bm{V}}{dt}d\omega \\
  &= \int_{V_p(t)} \bm{j}_p\times\bm{B}d\omega
    = 2\pi \int_S j_\varphi \grad (\Psi_w + \Psi_{\text{ext}}) d\bm{x}'
\end{align}
where Reynolds transport theorem and the continuity equation
$\partial_t\rho +\div\rho\bm{V}=0$ was used to pull the time
derivative inside the integral. Motion cannot arise from the plasma's
self-interaction, especially if it is detached from the wall
\cite{pustovitov-2015}, which is why only the wall and external coil
poloidal fluxes matter. This observation justifies a Taylor expansion
around the current centroid in order to write the vertical momentum
equation as
\begin{align}
  \frac{d P_z}{dt} = - l I_p\left[ B_{w+\text{ext},R}(\bm{x}_p) + \frac{1}{2}\bm{K}:\grad\grad B_{w+\text{ext},R}(\bm{x}_p) + \ldots\right]
\end{align}
where $l = 2\pi R$ is the length of the magnetic axis around the
torus. The first term is recognised as the force on a current-carrying
wire located at the current centroid. For the same reasons as for
equation (\ref{eq:psipatwall}), we neglect the second term. The rigid
body approximation and its interpretation is particularly
delicate. Its validity strongly depends on what happens in the open
field-line plasma or so-called halo region. Return flow patterns in
the halo region have been observed to facilitate flux redistribution
in resistive MHD simulations~\cite{aydemir}. These flows may however
be an artefact of treating the region beyond the last-closed
flux-surface as a cold low-density plasma instead of a true vacuum.
The separation between Alfvén and resistive timescales is usually
quite poor in numerical simulations such that inertia terms and
line-tying can become spuriously important and override the real
physical situation.
\section{Resistive wall described by multiple toroidal coils}
\label{sec:multiple_wires}
\amendment{Our analysis can be extended to more elaborate wall geometries by
using a Lagrangian principle in order to encode the inductive coupling
between multiple wall pieces and the moving plasma column. To include
resistive effects, the velocity gradient of the so-called Rayleigh
dissipation function is added to the Euler-Lagrange equations
\cite{essen-2009}. This framework extends the formulation of
engineering codes such as VALEN~\cite{valen}, commonly used to compute
wall Eddy currents for the design of vacuum vessels, to
self-consistently include the effect of a vertically displacing
plasma.}

Let us consider an arbitrary number of wires of thickness
$\Delta \ll a \ll l $, where $a$ is the characteristic distance
between the centre of the device and the components of the vacuum
vessel (minor radius) and $l=2\pi R$ is their length (major
radius). At $t=0$, the wire representing the plasma is centred at
$z(0)=0$ with a dynamical current $dY_p/dt=I_p(t)$, initially carrying
the plasma current $I_p(0)=I_0$. The vacuum vessel is represented by
multiple coils fixed at $z_i$. Their currents $dY_i/dt=I_i(t)$ are
dynamical quantities, initially set to $I_i(0)=0$. The Lagrangian is
written as
\begin{multline}
  \mathcal{L} = \frac{m}{2}\frac{dz}{dt}^2 + \frac{L_p}{2}\frac{dY_p}{dt}^2 \\
+\sum_i\left[ \frac{L_i}{2}\frac{dY_i}{dt}^2 +M(z-z_i)\frac{dY_p}{dt}\frac{dY_i}{dt}\right] \\
+ \sum_{i,j}\frac{M_{ij}}{2} \frac{dY_i}{dt}\frac{dY_j}{dt} + A_{\text{ext}}(z)\frac{dY_p}{dt}
\end{multline}
where $L_i$ is the self-inductance of each coil, $m$ the mass of the
plasma column, $M_{ij}=M(z_i-z_j)$ the mutual inductance between pairs
of coils and $A_{\text{ext}}(z)$ an external (driving) field (e.g. the
divertor coils).

The Rayleigh dissipation function that is added to the equations of
motion as a dissipative electromotive force is written as
\begin{equation}
\mathcal{D} = \frac{R_p}{2} \left(\frac{dY_p}{dt}\right)^2 + \sum_i \frac{R_i}{2}\left( \frac{dY_i}{dt}\right)^2
\end{equation}
where $R_i$ are the resistances of each wall wire.

The equations of motion are immediately found via
\begin{equation}
  \frac{d}{dt}\left(\frac{\partial \mathcal{L}}{\partial \dot{x}} \right) -\frac{\partial\mathcal{L}}{\partial x} + \frac{\partial \mathcal{D}}{\partial \dot{x}} = 0
\end{equation}
for $x \in \{ z,Y_p,Y_i\}$.  

It is convenient to use a similar normalisation as before
\begin{align}
  L_c &= \frac{\mu_0l}{2\pi} & 
  t_c &= \sqrt{ \frac{ m a^2}{L_c I_0^2}}
 = \frac{\sqrt{\mu_0 \rho_c}}{B_c}a   = \frac{a}{v_A}.
\end{align}
so that our variables become dimensionless
\begin{align}
\bar{z} &= z/a   &  \bar{I}_i &= \dot{y}_i= I_i/I_0   &   \tau &= t/t_c  ,
\end{align}
\begin{equation}
\begin{aligned}
\bar{\mathcal{L}} &= \mathcal{L} /L_cI_0^2 &
\bar{\mathcal{D}} &=  \mathcal{D} t_c/ L_cI_0^2 \\
\bar{L}_i  & = L_i/L_c &
\bar{M}(\bar{z}) &= M(z)/L_c  \\
\aext(\bar{z}) &=  A_{\text{ext}}(z) / L_c I_0 &
 \bar{R}_i &= R_i t_c/ L_c
\end{aligned}
\end{equation}
The normalised Lagrangian then reads
\begin{equation}
  \label{eq:norm_lagrangian}
  \bar{\mathcal{L}} =\tfrac{1}{2}\dot{\bar{z}}^2 + \tfrac{1}{2}\bar{L}_p \dot{y}_p^2 +
  \tfrac{1}{2}\dot{\vec{y}}\mathbb{M}\dot{\vec{y}} + \dot{y}_p \vec{M}\cdot \dot{\vec{y}} +
  \aext(\bar{z}) \dot{y}_p
\end{equation}
where $\dot{\bar{z}}=d\bar{z}/d\tau$,
$\mathbb{M}_{ij} = \bar{L}_i \delta_{ij} + \bar{M}_{ij} (\text{no
  sum})$ is the constant matrix of normalised wall
inductances\footnote{Explicitly,
  $\bar{M}_{ij}=\bar{M}(\bar{z}_i-\bar{z}_j)$ and $\bar{M}_{ii}=0$},
$\bar{M}_i = \bar{M}(\bar{z}-\bar{z}_i)$ is a vector of the normalised
plasma-wall mutual inductances and $\dot{y}_i = \bar{I}_i$ the vector
of normalised wall currents, $\dot{y}_p = \bar{I}_p$ the normalised
plasma current. A complementary description of the Lagrangian as well
as a general method to derive the inductance and resistance matrices
can be found in appendix \ref{sec:lagrangian}.

The Euler-Lagrange equations of motion become
\begin{align}
  \ddot{\bar{z}} &= \bar{I}_p\left(\vec{M}'\cdot \vec{I} + \aext'\right) \label{eqn:p_motion}\\
 \frac{d}{d\tau} \left[\bar{L}_p \bar{I}_p  + \vec{M}\cdot\vec{I} + \aext\right] &= -\bar{R}_p \bar{I}_p \label{eqn:flux_p}\\
 \frac{d}{d\tau}\left[\mathbb{M}\vec{I} + \bar{I}_p \vec{M} \right] &= -\mathbb{R}\vec{I} \label{eqn:flux_c}
\end{align}
where $\vec{M}' = d\vec{M}/d\bar{z}$
and $\mathbb{R}_{ij} = \bar{R}_i\delta_{ij} (\text{no sum})$ the diagonal
matrix of normalised wall resistances. The left-hand side of equations
(\ref{eqn:flux_p}-\ref{eqn:flux_c}) expresses the conservation of flux
through each wire, which is a direct consequence of the invariance of
the Lagrangian with respect to $y_p$ and $y_i$.

In the absence of resistive effects, the system is again integrable by
virtue of flux conservation and the dynamics can be resolved through
quadrature equations. A strong effective potential can be traced which
ensures that the system is non-linearly stable for plasma motion
occurring on Alfvénic times scales.

\subsection{Resistive decay of wall currents and characteristic plasma
  motion}
\label{sec:multiple_decay}
The slow drift of the equilibrium position $\bar{z}_e(\tau)$ is obtained
in the limit where the fast Alfénic dynamics have relaxed and force
balance between the induced wire currents and the external potential
is achieved, i.e. $\ddot{\bar{z}}_e \ll 1$. In this case, the vanishing
plasma acceleration provides a (scalar) constraint on the evolution of
the wall currents
\begin{equation}
  \ddot{\bar{z}} = \vec{M}'\cdot \vec{I} + \aext'=0.
\end{equation}
This equation suggests that the wall currents can be considered as a
field (instead of dynamical variables) satisfying
\begin{equation}
  \label{eq:field_current}
  \vec{I}(\bar{z}) = -\frac{\vec{M}'\aext'}{|\vec{M}'|^2} + \vec{I}_\perp(\bar{z})
\end{equation}
where $\vec{I}_\perp\cdot \vec{M}' = 0$ is an undetermined
orthogonal component. By differentiating the last two equations with
respect to $\bar{z}$, one shows that
\begin{equation}
  \vec{I}' = -\frac{\vec{M}'}{|\vec{M}'|^2}\aext'' +\aext'\left(2\frac{\vec{M}'}{|\vec{M}'|^2} \frac{\vec{M}''\cdot\vec{M}'}{|\vec{M}'|^2} -\frac{\vec{M}''}{|\vec{M}'|^2}\right)+ \vec{I}'_\perp
\end{equation}
and $\vec{I}'_\perp \cdot \vec{M}' + \vec{I}_\perp\cdot\vec{M}''=0$.

To avoid unnecessary complications, we will assume that the variation
of the plasma current is small enough not to interfere with the VDE
dynamics\footnote{The variation of plasma current can be included in
  the model presented. The treatment becomes more algebraic and
  further under-determined. The interesting limit where the wall is a
  perfectly conducting wall but the plasma current decays rapidly is
  discussed elsewhere~\cite{kiramov-breizman}.}, and use
$\bar{I}_p=1$. By projecting the vector equations for the wall
currents (\ref{eqn:flux_c}) onto $\vec{M}' \mathbb{R}^{-1}$, one then
obtains a single differential equation for the equilibrium position
\begin{align}
\label{eq:multiple_resistive_equation}
  \dot{\bar{z}} = \frac{\aext'}{\vec{M}'\mathbb{R}^{-1}\vec{M}' + \vec{M}'\mathbb{R}^{-1}\mathbb{M} \vec{I}'}
\end{align}
similarly as in (\ref{eq:resistive_equation}). If $\vec{I}'_\perp$ is
neglected, the ODE has no unknowns and can be treated
analytically. The component $\vec{I}'_\perp$ is supposedly driven by
the fast Alfvénic oscillations, that are damped away in less that one
wall time. The term $\vec{I}'_\perp$ is thus likely to play a minor
role in the slow relaxation process. Equation
(\ref{eq:multiple_resistive_equation}) is an extremely general
reduction of the VDE dynamics to a single ODE. It can be integrated
numerically with a time-step of the order of the wall time, even for
more elaborate wire models and geometries. Solving the full system of
circuit and motion equations (\ref{eqn:p_motion}-\ref{eqn:flux_c})
would require time-steps of the order of the Alfvén time.

\begin{figure}
  \begin{subfigure}[a]{\columnwidth}
    \includegraphics[width=\linewidth]{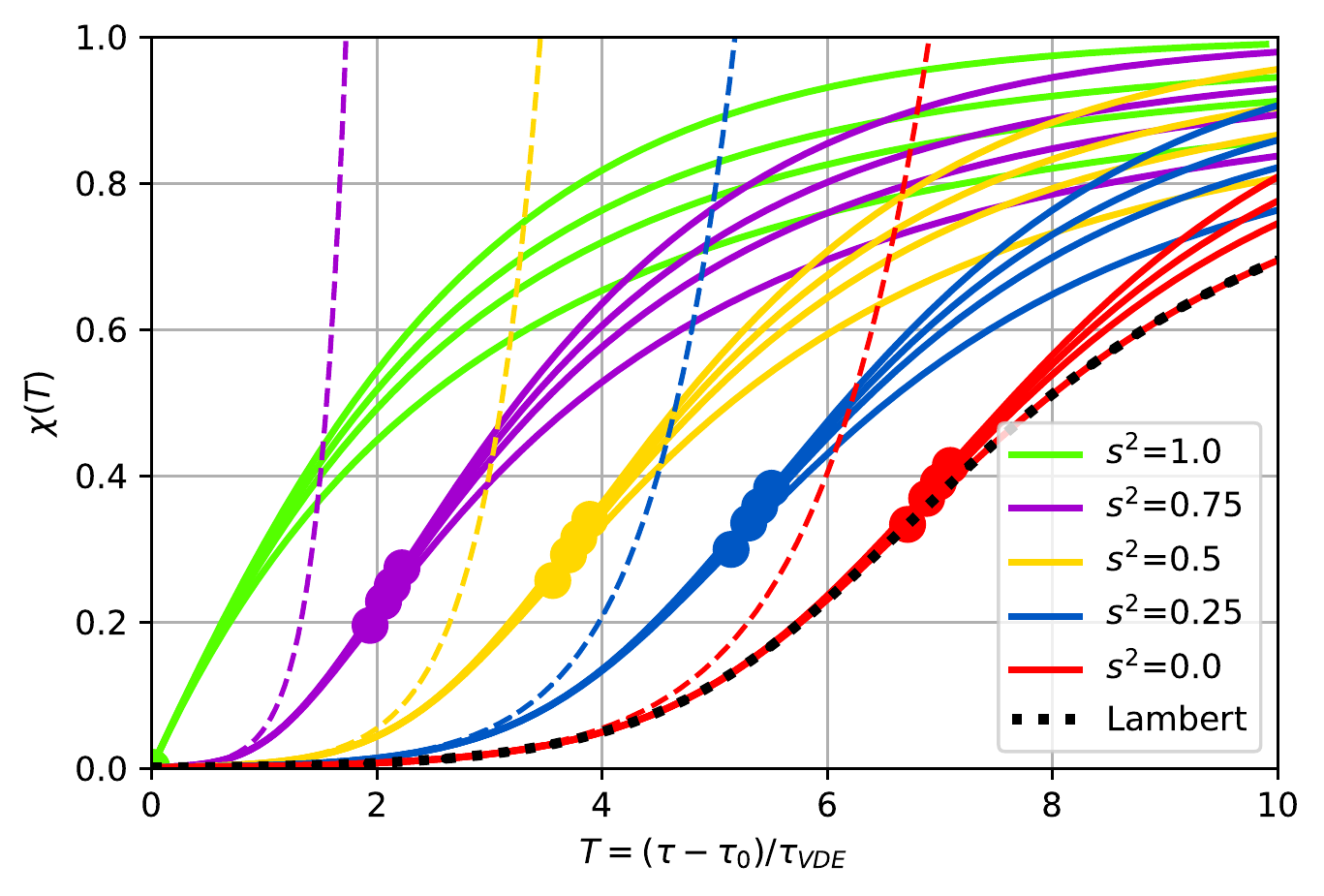}
    \caption{slowed down by a single wall wire.} %
    \label{fig:unstable_single}
  \end{subfigure}
\begin{subfigure}[b]{\columnwidth}
  \includegraphics[width=\linewidth]{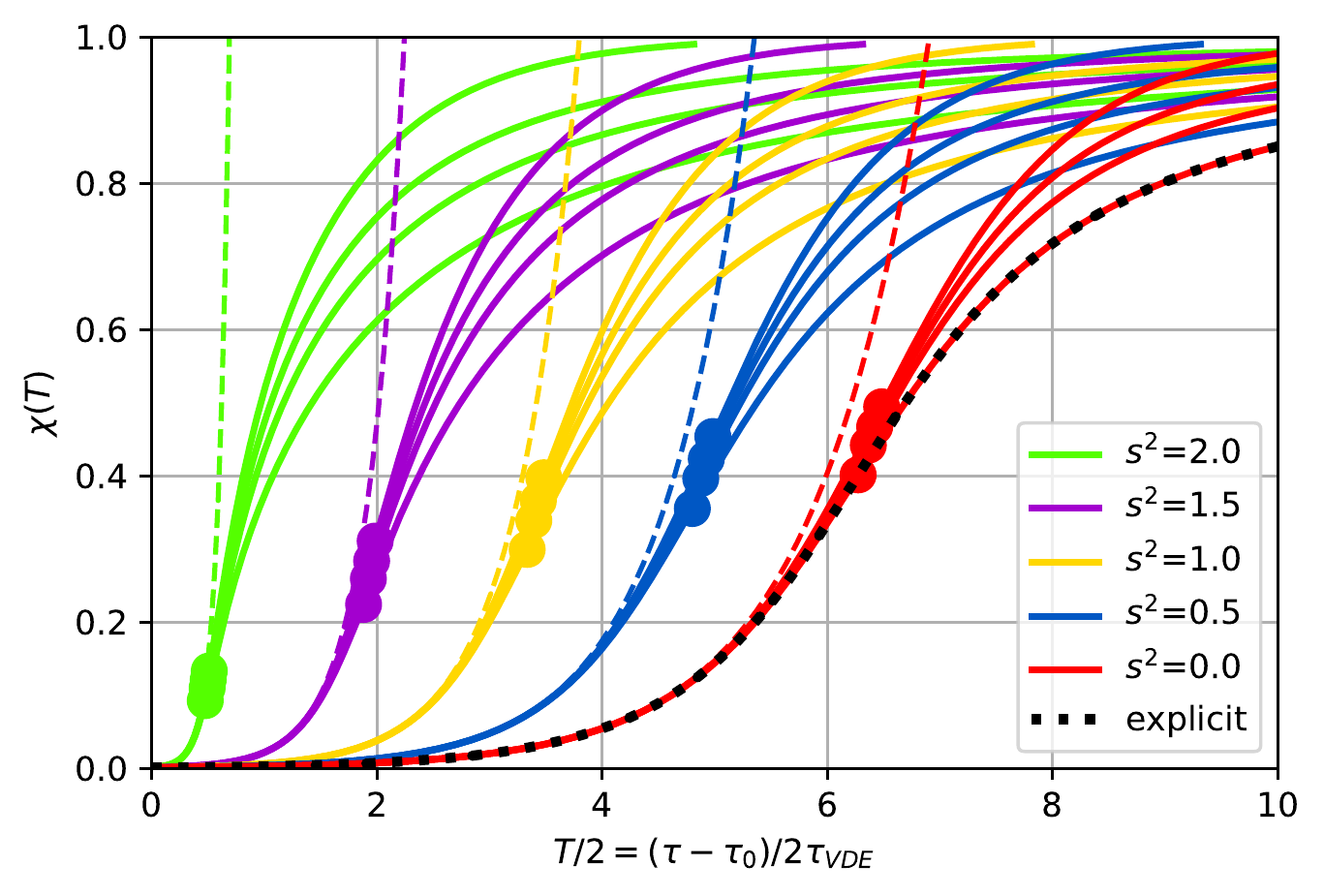}
    \caption{slowed down by two wall wires.}
    \label{fig:unstable_two}
  \end{subfigure}  
  \begin{subfigure}[c]{\columnwidth}
    \includegraphics[width=\linewidth]{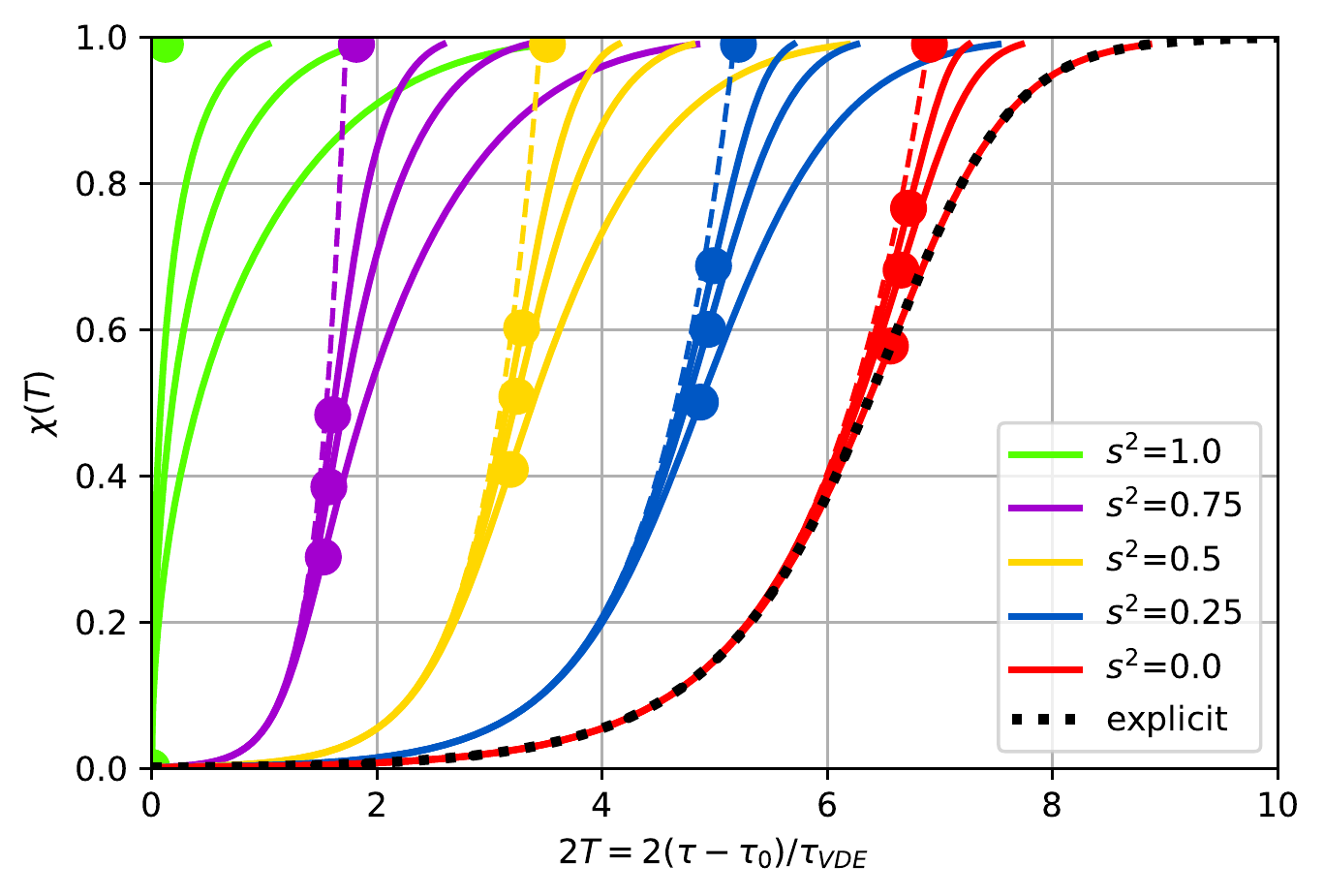}
    \caption{surrounded by a cylindrical conducting shell.}
    \label{fig:unstable_shell}
  \end{subfigure}
  \caption{Vertical drift of a plasma column due to the external field
    by two divertor coils (\ref{eq:divertor_field}) for varying values
    of $s^2=\gammavde/\gammalr$ (sets of curves with different colour,
    see legend) and $h^{-2} \in \{0,1/3,2/3,1\}$ (bottom to top curves
    of same colour). The initial exponential behaviour for
    $T\rightarrow - \infty$ is traced by the dashed colour lines. The
    inflexion points, $\ddot{\bar{z}}_*=0$ are denoted with coloured
    circles. The black dotted line represents the explicit solution
    $\bar{z}(T,h\rightarrow \infty, s^2\rightarrow 0)$. For better
    comparison, the x-axis is adjusted so that the span is identical
    for the three wall models.}
\end{figure}

As an illustrative example, we consider the wall to be represented by
two wires, located at $\bar{z}_u=1$ and $\bar{z}_d=-1$. The mutual
inductance between two circular coils of equal major radii is well
approximated by that of two parallel wires as discussed in appendix
\ref{sec:inductances}. We thus use $\bar{M}'(\bar{z}) = -1/\bar{z}$.
Assuming that the two wall coils have equal resistance $\bar{R}_w$ and
self-inductance $\bar{L}_w$, and are subject to the divertor field
(\ref{eq:divertor_field}), an implicit solution to
(\ref{eq:multiple_resistive_equation}) is obtained with the same
structure as (\ref{eq:divertor_single})
but with
\begin{align}
  \label{eq:two_wires}
  F(\bar{z}) =&   \frac{2\bar{z}^2}{1-\bar{z}^2} + \ln\left|\frac{2\bar{z}^2}{1-\bar{z}^2}\right|\\
  H(\bar{z}) =&\frac{2\bar{z}^2}{1-\bar{z}^2}+\ln(1-\bar{z}^2) \\
  K(\bar{z}) =&\tfrac{1}{2}\left(1-\frac{\bar{M}_{ud}}{\bar{L}_w}\right)\ln\left|\frac{2\bar{z}^2}{1+\bar{z}^2}\right| \nonumber\\
&+\ln\left|\frac{1-\bar{z}^2}{1-h^{-2}\bar{z}^2}\right|\nonumber\\
&+\frac{\bar{M}_{ud}}{\bar{L}_w}\left(\frac{1-h^{-2}}{1+h^{-2}}\ln\left|\frac{1+\bar{z}^2}{1-h^{-2}\bar{z}^2}\right|
+\frac{2\bar{z}^2}{1+\bar{z}^2}\right)
\end{align}
whose behaviour is shown on figure \ref{fig:unstable_two}) for various
values of $s^2$ and $h$. The initial growth rate in this configuration
is
  \begin{equation}
  \label{eq:init_growth_two}
  \frac{\dot{\bar{z}}}{\bar{z}} \overset{\bar{z}\rightarrow 0}{\longrightarrow} \frac{\gammavde}{2- s^2(1-\bar{M}_{ud}/\bar{L}_w)},
\end{equation}
where $\gammavde = \bar{R}_w \gammaext^2$, which is about half of what was obtained
with only one wall wire. This is expected because the combined
cross-section of the conducting parts has doubled and consequently the
total resistance (as a parallel circuit) reduced by a factor two. The
time axis of figure \ref{fig:unstable_two}) is scaled so that the
initial slopes match with figure \ref{fig:unstable_single}) as if the
total cross-sections are the same.

In the limit of far divertor coils, $h\rightarrow \infty$, the force
linearly increases from the centre point, $\aext' = \gammaext^2\bar{z}$. An
explicit solution of (\ref{eq:two_wires}) for $s^2\rightarrow 0$ is
obtained
\begin{equation}
  \label{eqn:two_wires_lin}
  \bar{z}(T;h\rightarrow \infty, s^2\rightarrow 0) = \frac{1}{\sqrt{1+2/W(e^T)}}
\end{equation}
where $T=\gammavde(\tau-\tau_0)$. This curve is depicted by the dashed
curve on figure \ref{fig:unstable_two}).

As before, the furthest inflexion point is found in the extreme case
where the divertor coils are touching the wall, i.e. $h\rightarrow 1$
and the driving force is small with respect to the Eddy potential,
$s^2\rightarrow 0$.
Inflexion is then found at
$\bar{z}_*(h\rightarrow 1,s^2\rightarrow 0) = \sqrt{\sqrt{5}-2}\cong
48.6\%$ across the vacuum vessel, which is again a purely geometric
result. The inflexion point is brought closer to the centre if the
external drive is increased.
\section{Resistive wall as a cylindrical shell}
\label{sec:cylindrical_shell}
\amendment{The plasma is not simply bounded by toroidal wires, it is
(topologically) fully enclosed by a conducting vessel. It is thus
important to study the currents that form within a thin cylindrical
shell as the plasma column moves in its interior. The added dimension
in 2D shells versus 1D wires allows for extra diffusion of magnetic
flux in the poloidal direction, which affects the speed and behaviour
of the VDE.}

\begin{figure}
  \centering
  \includegraphics[width=0.5\linewidth]{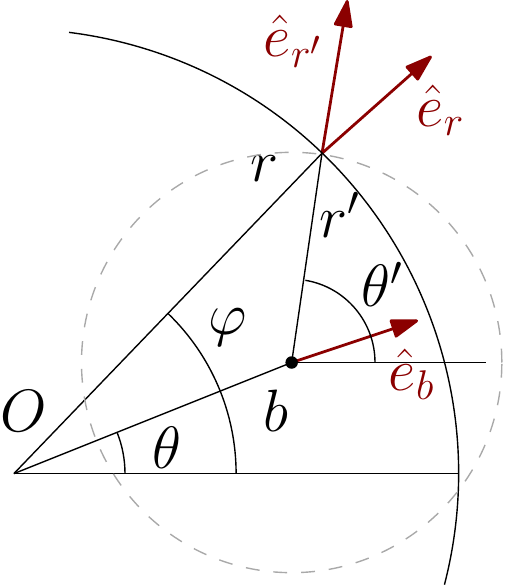}  
  \caption{Choice of cylindrical coordinates and definitions of angles and radii.}
  \label{fig:shell_scheme}
\end{figure}
In what follows, the origin of the coordinate system coincides with
the axis of symmetry of the shell. The radius of the shell is the
minor radius $a$ and its thickness is $\Delta$. The plasma is modelled
as a straight wire floating at $x(t)=b(t)\cos[\theta(t)]$ and
$y(t)=b(t)\sin[\theta(t)]$. The choice of coordinates and definition
of angles and distances are sketched on figure \ref{fig:shell_scheme}.
\subsection{Fields from plasma wire and shell surface currents}
First, the field produced by the plasma wire is expressed in
cylindrical coordinates. As shown in the appendix
\ref{sec:potential_wire}, this can be performed either by the
multipole expansion of the vector potential or by solving the 2D
Laplace equation given the plasma current density
\begin{align}
  \bm{j}_p 
&= I_p\delta(r-b)\delta(\varphi - \theta)\grad r\times \grad \varphi
\end{align}
The resulting potential functions representing the plasma magnetic
field $\bm{B}_p =\grad \Phi_p = \grad\Psi_p\times\grad z$ are
\begin{align}
 \Psi_p(r>b,\varphi) &= \frac{\mu_0I_p}{2\pi}\left[-\ln\frac{r}{b} +\sum_{m=1}^\infty \left(\frac{b}{r}\right)^m  \frac{\cos[m(\varphi-\theta)]}{m}\right] 
\end{align}
and
\begin{align}
  \Phi_p(r>b,\varphi) &= \frac{\mu_0 I_p}{2\pi}\left\{\varphi + \sum_{m=1}^\infty \left(\frac{b}{r}\right)^m  \frac{\sin[m(\varphi-\theta)]}{m}\right\}
\end{align}

Second, the surface current density flowing in the $z$-direction on
the thin shell at $r=a$ is represented as
\begin{equation}
  \bm{j}_s(r,\varphi,t) = \delta(r-a) \kappa'(\varphi,t) \grad r \times \grad \varphi = \delta(r-a) \grad r\times \grad \kappa
\end{equation}
where the surface current function $\kappa'(\varphi,t)$ is expanded in
a Fourier series as
\begin{align}
  \kappa'(\varphi,t) &= \sum_{m=0}^\infty\left[ j^s_m(t)  \sin(m\varphi) + j^c_m(t)\cos(m\varphi)\right]\nonumber\\
& = \frac{I_0(t)}{2 \pi} + \sum_{m=1}^\infty \frac{I_m(t) }{\pi} \cos[m(\varphi- \alpha_m(t))]
\label{eq:kappa_def}
\end{align}
By the same exercise as in appendix \ref{sec:current_derivation}, the
associated potential function in the region enclosed by the shell at
$r=a$ is found to be
\begin{align}
  \Phi_s(r<a,\varphi,t) &= -\frac{\mu_0 }{2\pi}\sum_{m=1}^\infty I_m \left(\frac{r}{a}\right)^m \frac{\sin[m(\varphi-\alpha_m)]}{m} 
\end{align}
and therefore the poloidal flux function is deduced by the
Cauchy-Riemann condition (\ref{eq:cauchy-riemann}), 
\begin{align}
  \Psi_s(r<a,\varphi,t) &= \frac{\mu_0 }{2\pi}\sum_{m=1}^\infty I_m \left(\frac{r}{a}\right)^m\frac{\cos[m(\varphi-\alpha_m)]}{m}.
 \label{eq:flux_shell}
\end{align}
One can express the magnetic field generated by surface currents on
the cylindrical shell anywhere between $b<r<a$ via
$\bm{B}_s = \grad \Phi_s = \grad\Psi_s\times\grad z$.

\subsection{Circuit and motion equations}
If only interested in the (radial) motion of the plasma column, the
phase $\alpha_m(t) = \theta(t) = 0$ can be held fixed. The initial value
of $I_p(0)$ is used to normalise the currents, $\bar{I}_p(t)=I_p/I_p(0)$ and
$\bar{I}_m(t) = I_m/I_p(0)$.

The azimutal component of the magnetic field from the shell produces
the radial force on the wire
\begin{align}
M \frac{d^2b}{dt^2} &= \bm{F}\cdot\grad r\Big|_{b,\theta} 
= l I_p \grad z \times \bm{B}_s \cdot\grad r + F_{ext}\\
&= \frac{ \mu_0 l I_p}{2\pi b} 
\sum_{m=1}^\infty I_m \left(\frac{b}{a}\right)^m  + F_{ext}
\end{align}
where $\cos[m(\theta-\alpha_m)] =1$ was used, $M$ denotes here the
plasma mass and $F_{ext}=I_p dA_{ext}/dz$ is the external force from
the divertor fields. In normalised units, the equation of motion of
the plasma column reads
\begin{align}
  \label{eq:shell_eom}
 \ddot{\bar{z}}  = \bar{I}_p\left[\sum_{m=1}^\infty \bar{I}_m \bar{z}^{m-1} + \aext'(\bar{z})\right]
\end{align}
where $\bar{z}(\tau) =b/a$, $\tau=t/t_c$, $t_c = \sqrt{M a^2/L_cI_p^2}$,
$L_c=\mu_0 l/2\pi$ as before. This equation complies with
(\ref{eqn:p_motion}).

In the vacuum between the shell and the plasma column, the total
magnetic field produced by the plasma current $I_p(t)$ and surface
currents $I_m(t)$ is 
\begin{multline}
  \bm{B}(b<r<a,\varphi,t) = \frac{\mu_0}{2\pi} \grad \Bigg\{I_p \varphi \\
+ \sum_{m=1}^\infty\frac{\sin(m\varphi)}{m} \Bigg[ I_p\left(\frac{b}{r}\right)^m- I_m \left(\frac{r}{a}\right)^m\Bigg]\Bigg\}
\end{multline}
where the phase $\alpha_m(t)=0$ was omitted.

The time-variation of the total magnetic field induces an electric
field (Faraday's law), $\partial_t \bm{B} + \curl\bm{E} = 0$, which
drives current within the thin conducting shell via Ohm's law, $\bm{E}
= \eta_w \bm{j}$.
The normal component of the magnetic field is the only useful
projection since it is continuous within the shell. In the limit of an
infinitesimally thin wall, it can be assumed to be radially constant
across the shell. We thus have
\begin{equation}
 \partial_\tau\left(\frac{\bm{B}}{B_c}\cdot\hat{\bm{e}}_r\right )_a= -\sum_{m=1}^\infty \sin(m\varphi)
  \frac{d}{d\tau}\left[\bar{z}^m \bar{I}_p + \bar{I}_m\right]
\label{eq:brad}
\end{equation}
where $B_c = \mu_0 I_p(0)/2\pi a$ is the initial plasma poloidal
field at the shell.

The current distribution inside the thin shell is also assumed to be
radially constant, with the following representation of the delta
function \cite{boozer-2015}
\begin{equation}
  \delta(r-a) \overset{\Delta\rightarrow 0}{\longleftarrow}
  \begin{cases}
    \frac{1}{\Delta} & a-\Delta/2< r < a+\Delta/2 \\
    0 & \text{elsewhere}
  \end{cases}
\end{equation}
so that, at $r=a$, we have
\begin{align}
\hat{\bm{e}}_r\cdot \curl \bm{E}\big|_a &= \grad r\cdot \curl (\eta_w \bm{j})\Big|_a \nonumber \\
&= -\frac{\eta_w}{\Delta} \div [\kappa'(\varphi,t) \grad r\times (\grad r\times\grad \varphi)]\Big|_a \nonumber\\ 
& = \frac{\eta_w}{\Delta}\div ( \kappa'\grad\varphi)\Big|_a = \frac{\eta_w}{\Delta} \frac{\kappa''}{a^2} 
\label{eq:kappa_curl}
\end{align}
where $\kappa'' = \partial_\varphi \kappa'$, $(\grad \varphi)^2=1/r^2$  and $\grad^2\varphi= 0$. 

Using the representation for $\kappa$ of (\ref{eq:kappa_def}) and
combining (\ref{eq:brad}) and (\ref{eq:kappa_curl}) in Faraday's and
Ohm's law one obtains a list of decoupled circuit equations. In
normalised variables, each mode $m\geq 1$ obeys
\begin{align}
  \label{eq:shell_cflux}
  \frac{d}{d\tau}\left[ \frac{1}{m}\bar{I}_m + \bar{I}_p\frac{\bar{z}^m}{m}\right] &= - 2 \bar{R}_w \bar{I}_m
\end{align}
where $\bar{R}_w^{-1} = S_w (\Delta /a)$, $S_w = \mu_0 a v_A/\eta_w$ is the
wall Lundquist number and $v_A = a/t_c$ the characteristic Alfvén
speed as before. This infinite list of equations is the spectral
version of (\ref{eqn:flux_c}).

The last circuit equation originates from the flux change through the
plasma wire. Upon evaluating $\partial_t \Psi_s|_{b,\theta}$ from
(\ref{eq:flux_shell}), one arrives at
\begin{align}
  \frac{d}{d\tau} \left[\bar{L}_p \bar{I}_p + \sum_{i=1}^\infty \frac{\bar{z}^m}{m} \bar{I}_m+  \aext\right] = -\bar{R}_p \bar{I}_p
  \label{eq:shell_pflux}
\end{align}
which is analogous to (\ref{eqn:flux_p}). Notice that the shell cannot
produce $m=0$ flux within its interior and therefore cannot oppose to
the decay of plasma current (Gauss law\footnote{In the limit of
  perfectly conducting shell, the normal component of the magnetic
  field to the wall is frozen, but the $m=0$ component of the magnetic
  field is purely tangential.}).

In comparison with the model with multiple wires, the shell displays a
diagonal self-inductance matrix $\mathbb{M}_{mn} = \delta_{mn}/m$ and
scalar resistance $\mathbb{R}_{mn} = 2\bar{R}_w \delta_{mn}$. The mutual
inductance vector is identified as $\bar{M}_m = \bar{z}^m/m$ and we remark
that
\begin{align}
  \vec{M}'\cdot\vec{M}' &= \frac{1}{\bar{z}^2}\sum_{m=1}^\infty \bar{z}^{2m} = \frac{1}{1-\bar{z}^2}\\
  \vec{M}'\mathbb{M}\vec{M}'& = \frac{1}{\bar{z}^2}\sum_{m=1}^\infty\frac{\bar{z}^{2m}}{m} =  -\frac{1}{\bar{z}^2}\ln|1-\bar{z}^2|
\end{align}

In the limit of a perfectly conducting shell where $\bar{R}_w\rightarrow 0$,
the system is integrable by virtue of flux conservation. 
The plasma is stabilised by the presence a diverging potential, which
can be shown to be $V_{\text{eff}} = -\frac{1}{2}\ln|1-\bar{z}^2|$. %
\subsection{Resistive decay of surface currents and characteristic
  plasma motion}
\label{sec:resdecay_shell}
In the resistive decay regime where the slow evolution of the
equilibrium position is sought for, the assumption of time-scale
separation, $\ddot{\bar{z}}_e\ll 1$, suggests that the surface
currents satisfy the spectral analogue of relation
(\ref{eq:field_current}). Force balance requires that the mode
amplitudes be the fields
\begin{align}
  \bar{I}_m(\bar{z}) =  -\bar{z}^{m-1}(1-\bar{z}^2)\aext'
\end{align}
Assuming that $\bar{I}_p\sim 1$ for simplicity, the motion of the equilibrium
position reads from (\ref{eq:multiple_resistive_equation})
\begin{align}
  \dot{\bar{z}} = \frac{2\bar{R}_w\aext'}{
  \frac{1}{1-\bar{z}^2}
  - \frac{\aext'}{\bar{z}}\left(1+\frac{1+\bar{z}^2}{\bar{z}^2}\ln|1-\bar{z}^2|\right)
  + \aext''\frac{1-\bar{z}^2}{\bar{z}^2}\ln|1-\bar{z}^2|}
  \label{eqn:cyl_resdyn}
\end{align}
%
which becomes for the specific divertor field of
(\ref{eq:divertor_field})
\begin{align}
  \dot{\bar{z}}\left\{\frac{1-h^{-2}\bar{z}^2}{(1-\bar{z}^2)\bar{z}}
- s^2\left[\frac{1}{\bar{z}} + (1-h^{-2})\frac{2}{\bar{z}}\frac{\ln|1-\bar{z}^2|}{1-h^{-2}\bar{z}^2}\right]
  \right\} = 2\gammavde
\end{align}
where $\gammavde = \bar{R}_w \gammaext^2$ as before. The implicit solution of the
above ODE in the form of (\ref{eq:divertor_single}) is
\begin{align}
  F(\bar{z}) =& \frac{1}{2}\ln\left|\frac{\bar{z}^2}{1-\bar{z}^2}\right| \\
  H(\bar{z}) =& \frac{1}{2}\ln\left|\frac{1}{1-\bar{z}^2}\right| \\
  K(\bar{z}) =& \ln\bar{z} - (1-h^{-2})
            \left\{ \litwo(\bar{z}^2) + \litwo\left[-\frac{h^{-2}(1-\bar{z}^2)}{1-h^{-2}}\right]\right. \nonumber\\
  & \left.+ \ln|1-\bar{z}^2|\ln\left|\frac{1-h^{-2}\bar{z}^2}{1-h^{-2}}\right| \right\}
\end{align}
where $\litwo(z) = -\int^z_0 \frac{\ln(1-u)}{u} du $ is the so-called
Spence function or dilogarithm.
%
%
%
This solution is displayed on figure \ref{fig:unstable_shell}) for
various values of $s^2$ and $h$. The initial growth rate in this
configuration is
\begin{equation}
  \label{eq:init_growth_shell}
  \frac{\dot{\bar{z}}}{\bar{z}}\overset{\bar{z}\rightarrow 0}{\longrightarrow}
  \frac{2\gammavde}{1-s^2}
\end{equation}
which is twice as fast as for wire models at equal conducting
cross-sections. This is an interesting difference between wire and
shell models, where the extra dimension allows for diffusion of flux
across the surface, additionally to in the direction of current. The
time axis of figure \ref{fig:unstable_shell}) is scaled so that the
initial slopes match with figure \ref{fig:unstable_single}) for easier
comparison.

In the limit where the divertor coils are far from the wall,
$h \rightarrow \infty$, an explicit solution for $s^2\rightarrow 0$ is
given by
\begin{equation}
\bar{z}(T;h\rightarrow \infty,s^2\rightarrow 0) = \frac{1}{\sqrt{1 + e^{-4T}}}
\end{equation}
This behaviour is illustrated on figure \ref{fig:unstable_shell}) by
the dashed black line.

In the extreme case where the divertor coils are radially located on
the shell, $h \rightarrow 1$, the solution tends to preserve the
asymptotic exponential behaviour
\begin{equation}
  \bar{z}(T;h\rightarrow 1,s^2) = e^{2T/(1-s^2)}
\end{equation}
and inflexion never occurs, as depicted by the dashed colour curve on
figure \ref{fig:unstable_shell}).
\section{Conclusion}
\label{sec:conclusion}
Assuming that the plasma column behaves as a rigid rod during a VDE, a
series of analytic expressions for its non-linear vertical drift
across the vacuum vessel were derived. First, a basic model where the
wall is treated as a single conducting wire was used to identify four
timescales: the fast Alfvén, the fast oscillations around the minimum
of effective potential, the slow current decay (``L over R'') time and
the slower VDE rate. Well separated timescales between the Alfvén
dynamics and VDE evolution was assumed based on experimental
evidence. The linear analysis revealed two oscillatory modes and a
slow exponentially growing mode coinciding with the vertical
instability. The non-linear dynamics were shown to be solvable in the
limit where the wall is a perfect conductor and the induced currents
fully stabilise the plasma column by generating a strong effective
potential. With weak wall resistivity, the equilibrium point relaxes
towards the vessel at the rate of current dissipation. In this regime,
the induced wall currents compensate the divertor field at the
position of the plasma column. The force balance condition was used to
eliminate the fast oscillatory motion and capture the non-linear
relaxation process into a single ODE.

The model was extended to an arbitrary number of wall wires via a
Lagrangian principle, where the decay of currents is incorporated
through the Rayleigh dissipation function. A general ODE for the
vertical drift of the plasma, equation
(\ref{eq:multiple_resistive_equation}), was derived in the limit of
weak resistivity. The case where the plasma is surrounded by two wall
wires was studied analytically. Although the initial growth rate is
almost identical for equivalent combined cross-sections, the
characteristic motion with two wall wires was shown to be steeper and
the inflexion point farther than in the case with only a single wall
wire.

The methodology was finally applied to the case where the wall is
represented as a thin cylindrical shell. The vector potential produced
by surface currents on the wall was expressed in cylindrical
coordinates via the standard solution to the two-dimensional Laplace
equation. An infinite system of circuit equations was then obtained
for the Fourier modes of the surface current and the corresponding
resistance and inductance matrices of the shell were
identified. Inserted into the single ODE
(\ref{eq:multiple_resistive_equation}), the infinite series produced
logarithmic functions for the VDE dynamics and an analytic solution
was obtained.

A study of more general geometries for the surrounding conducting
structures can be performed analytically or numerically in a
straightforward, robust and efficient way by computing the wall
inductance and resistance matrices as well as the plasma-wall
mutual-inductance vector with the identities reported in appendix
\ref{sec:lagrangian}.

\amendment{In all systems studied, it was found that the motion deviates from the
exponential growth expected from linear analysis and becomes a more
complex algebraic deceleration. It was observed that the instantaneous
growth rate, $\dot{\bar{z}}/\bar{z}$, monotonically decreases to zero
as the plasma reaches the wall, while the acceleration
$\ddot{\bar{z}}$ crosses zero at the inflexion point. For a given
divertor height, the inflexion point is closer to the wall when the
external field is weaker. The maximum inflexion point is a purely
geometric quantity that can be used experimentally as a reference
point to locate a change of physics. Indeed, experimental
reconstruction of the magnetic axis position often show that the
plasma column accelerates towards the wall near the end of a VDE. The
fact that our model contradicts this observation leads us to conclude
that, if no inflexion occurs, the physics controlling the VDE
evolution is non-inductive. If the magnetic axis is found to
accelerate, for example as in the NSTX \cite{gerhardt-2012}, the VDE
is most probably driven by the sharing of current between the edge
plasma and the wall, the scraping of the last-closed flux surfaces
and/or from the internal thermal quench dynamics. If the plasma slows
down in the vicinity of the wall, as seen in JET upward VDEs
\cite{gerasimov-2015}, the inductive component is dominant. In the
latter case, the picture of a slowly evolving three-dimensional
equilibrium \cite{zakharov-2008} seems a suitable model, while in the
former more physics (MHD and beyond) must be invoked.}
\section*{Acknowledgements}
The authors would like to acknowledge stimulating discussions with
S.C. Jardin, N. Ferraro, A. Boozer, J. Bialek and L. Zakharov. The
authors were supported by the Department of Energy Contract
No. DE-AC02-09CH11466 during the course of this work.

%
%
\section*{References}
\bibliographystyle{phjcp}
\bibliography{biblio}
\appendix
\section{Circuit equations, inductance and resistance matrices from variational principle}
\label{sec:lagrangian}
The Lagrangian (\ref{eq:norm_lagrangian}), which leads to the motion
and circuit equations (\ref{eqn:p_motion}-\ref{eqn:flux_c}),
essentially consists of the plasma kinetic energy and the total
magnetic energy produced by the plasma, wall and external coils. The
framework is valid for 2D and 3D conducting structures, beyond the
application to discrete set of wires and filaments. In fact, the
equations for the plasma surrounded by a cylindrical shell
(\ref{eq:shell_eom}, \ref{eq:shell_cflux} and \ref{eq:shell_pflux})
could have been derived using the Lagrangian principle
directly. Indeed, recalling that
$\div(\bm{A}\times\bm{B}) =(\curl \bm{A})\cdot \bm{B} -
\bm{A}\cdot(\curl\bm{B})$ and assuming that the magnetic field
vanishes at infinity, the total magnetic energy can be expressed as
\begin{align}
  \mathcal{L}_{M}  &= \frac{1}{2\mu_0}\int_{\mathbb{R}^3} B^2 d\omega
 = \frac{1}{2}\int_\Omega \bm{A}\cdot \bm{j} d \omega
\end{align}
where $\Omega$ is the support of the current density
$\bm{j} =\curl\bm{B}/\mu_0$. Decomposing the total magnetic field into
the plasma, wall and external components
$ \bm{B} = \bm{B}_p + \bm{B}_w + \bm{B}_{ext} $, the wall inductance
matrix is obtained as
\begin{align}
N  \mathbb{M}_{mn} =\frac{1}{2}\frac{\partial^2}{\partial \bar{I}_m\partial \bar{I}_n} \int_{\text{wall}} \bm{A}_w \cdot\bm{j}_w d\omega
\end{align}
where $\bar{I}_m$ are adimensional degrees of freedom chosen to represent
wall currents and $N = L_cI_p(0)^2$ the normalisation factor. The
plasma-wall mutual inductance vector is
\begin{align}
N  \vec{M} = \frac{\partial^2}{\partial \vec{I} \partial \bar{I}_p}\int_{\text{wall}} \bm{A}_p \cdot\bm{j}_w d\omega 
 = \frac{\partial^2}{\partial \vec{I} \partial \bar{I}_p} \int_{\text{plasma}} \bm{A}_w \cdot\bm{j}_p d\omega 
\end{align}
where $\bar{I}_p = I_p/I_p(0)$ is the normalised plasma current. Notice the
reciprocity of the latter expression. Similarly, the external
potential generated by the divertor coils can be expressed as
\begin{align}
L_c I_p(0) \aext = \frac{\partial}{\partial \bar{I}_p} \int_{\text{plasma}} \bm{A}_{ext} \cdot\bm{j}_p d\omega. 
\end{align}

The Rayleigh dissipation function essentially conveys the Ohmic power
loss
\begin{align}
  \mathcal{D} = \frac{1}{2}\int_\Omega \eta \bm{j}^2 d\omega  
\end{align}
so that the resistance matrix may be expressed as
\begin{align}
\frac{ v_AN}{a}  \mathbb{R}_{mn} =\frac{1}{2}\frac{\partial^2}{\partial \bar{I}_m \partial \bar{I}_n} \int_{\text{wall}} \eta_w \bm{j}^2_w d\omega
\end{align}
where care must be taken with squared delta-functions from the
filament or surface current density.

The reader may verify that the diagonal inductance and resistance
matrices for the cylindrical shell presented in section
(\ref{sec:cylindrical_shell}) are correctly recovered through these
identities. They are particularly useful in the context of a spectral
or Finite Element representation of wall currents. It is mentioned
that the motion of the plasma column within a toroidally shaped wall
can be studied in this way too.
\section{Coil inductances}
\label{sec:inductances}
Self-inductance and mutual inductance of filaments are purely
geometric coefficients that can be acquired experimentally or
estimated analytically. The expression and interpretation of
self-inductance via Neumann formula is somewhat subtle because it
formally diverges \citep{rosa}. The normalised self-inductance of a
thin circular coil of perimeter $l=2\pi R$ is given for a uniform
current density by \cite{essen-2013}
\begin{equation}
\label{eq:self-inductance}
\bar{L} =\frac{L}{L_c} =  -\left[\ln\left(\frac{\epsilon \delta}{8}\right) + \frac{7}{4}\right]
\end{equation}
where $\delta = \Delta/a$ is the normalised diameter of the coils,
$\epsilon = a/R$ the inverse aspect ratio and
$L_c=\mu_0 l /2\pi = \mu_0 R$.


The normalised mutual inductance, $\bar{M}(\bar{z}) = M(z)/L_c$ where
$L_c = \mu_0 l/2\pi$, of two circular coils of equal major radius
$R = a/\epsilon$ is given as a function of the separating normalised
vertical distance $\bar{z} = z/a$ by \cite{essen-2009}
\begin{equation}
\bar{M}(\bar{z}) =  \frac{1}{k}\left[(2-k^2)K(k^2) - 2E(k^2)\right] 
\end{equation}
where $K(m) = \int_0^{\pi/2}\frac{d\theta}{\sqrt{1-m\sin^2\theta}}$ is
the complete elliptic integral of the first kind,
$E(m) = \int_0^{\pi/2}\sqrt{1-m \sin^2\theta}d\theta$ the complete
elliptic integral of second kind,
\begin{equation}
 k(\bar{z}) = \frac{1}{\sqrt{1 + (\frac{\epsilon \bar{z}}{2})^2}}
\end{equation}
and $\epsilon = a/R$ is the inverse aspect ratio.

It is noted that the argument of the elliptic functions is bounded by
$2/\sqrt{5}\cong 0.89 \leq k \leq 1$. Due to the logarithmic
divergence of $K(k)$ as $k\rightarrow 1$, the mutual inductance of two
circular coils becomes well approximated by
\begin{equation}
  \bar{M}(\bar{z}) \longrightarrow -\ln(\epsilon \bar{z})
\end{equation}
corresponding to the mutual inductance of two straight
wires. Toroidicity is seen to play a minor role in the derivative of
the mutual inductance representing the Lorentz force between two
wires, $\bar{M}'\rightarrow -1/\bar{z}$.
\section{External potential from divertor coils}
\label{sec:divpot}
Vertical Displacement Events (VDEs) are driven by the external
potential produced by the two divertor coils giving the plasma its
elongation. We assume for simplicity that the currents in each
divertor coil are fixed and identical $\bar{I}_c = I_c/I_p(0)$ and that the
coils are positioned symmetrically about the origin, at $z=z_c > a$
and $z=-z_c<-a$. The (normalised) external force is given by (see
appendix \ref{sec:inductances})
\begin{align}
  \label{eq:divertor_field}
  \fext& = \bar{I}_c [\bar{M}'(\bar{z}-h) + \bar{M}'(\bar{z}+h)] \\
  &= \bar{I}_c\left(\frac{1}{h-\bar{z}} -\frac{1}{h+\bar{z}}\right)
  = \frac{\gammaext^2\bar{z}}{1-h^{-2}\bar{z}^2}
\end{align}
where $\bar{z}=z/a$, $h=z_c/a$ and $\gammaext^2 \equiv 2 \bar{I}_c/h^2$ is the inverse
width of the quadratic potential at the unstable equilibrium point. In
toroidal shaped plasmas, one can invoke radial force balance and
$\curl \bm{B}_{ext}= 0$ in order to relate the $\gammaext$ parameter to the
so-called field decay index $n$ (parametrisation of the vertical field
as $ B_Z \propto R^{-n}$) and the plasma inductance
$\bar{L}_p =L_p/L_c$
\cite{lazarus-1990,shafranov-1963}
\begin{equation}
  \gammaext^2 \sim \epsilon^2\bar{L}_p  n
\end{equation}
where $\epsilon=a/R$ is the inverse aspect ratio.
\section{Field from the plasma wire on the cylindrical shell}
\label{sec:potential_wire}

In the vacuum region, the poloidal magnetic field produced by the
plasma wire satisfies two conditions. First, $\div\bm{B}_p=0$, true
for any magnetic field, implies that $\bm{B}_p=\curl\bm{A}_p$. By
symmetry along the cylinder's axis $z$, the vector potential is
represented by the poloidal flux function only as
$\bm{A}_p = \Psi_p\grad z$ or $\bm{B}_p = \grad \Psi_p\times \grad
z$. Secondly, $\curl\bm{B}_p = 0$ since there are no currents in the
vacuum region, which implies that the poloidal field is also
represented by the gradient of a potential function $\Phi_p$,
$\bm{B}_p = \grad \Phi_p$. By combining those conditions, one
concludes that both potentials function are harmonic,
$ \grad^2 \Phi_p = 0$ and $\grad^2 \Psi_p = 0$. They in fact form what
is called a pair of harmonic conjugate functions which satisfies
$\grad \Phi_p \cdot\grad\Psi_p = 0$ and Cauchy-Riemann equations
\begin{align}
  r\partial_r \Phi_p &= \partial_\varphi \Psi_p &
 \partial_\varphi \Phi_p &= -r\partial_r \Psi_p.
\label{eq:cauchy-riemann}
\end{align}
There are two ways to evaluate $\Psi_p$ and $\Phi_p$ in the natural
coordinates of the cylindrical shell. It is instructive to detail both
of them.
\subsection{Multipole expansion of the vector potential}
The magnetic field produced by a straight wire traversed by current
$I_p$ is easy to express as a function of the separating distance $r'$ 
as in figure \ref{fig:shell_scheme}
\begin{equation}
  \bm{B}_p(r',\theta') = \frac{\mu_0 I_p}{2\pi r'}\hat{\bm{e}}_{\theta'} \label{eqn:bnat}
\end{equation}
where $\hat{\bm{e}}_{\theta'}=\hat{\bm{e}}_z\times \hat{\bm{e}}_{r'} =
\grad z\times \grad r'$. The corresponding poloidal flux is then
written as
\begin{equation*}
  \Psi_p(r',\theta') = -\frac{\mu_0 I_p}{2\pi}\ln r'
\end{equation*}

For the purpose of calculating the effect of this field on a
cylindrical conducting shell, a change of coordinates is performed
\begin{align*}
  r \hat{\bm{e}}_r &= r' \hat{\bm{e}}_{r'} + b \hat{\bm{e}}_b &
  r' &= \sqrt{ r^2 + b^2 - 2 b r \hat{\bm{e}}_b\cdot\hat{\bm{e}}_r}
\end{align*}
where $\hat{\bm{e}}_b\cdot\hat{\bm{e}}_r = \cos(\varphi-\theta)$. 
The poloidal flux then reads
\begin{equation}
 \Psi_p(r,\varphi) = -\frac{\mu_0 I_p}{4\pi} \ln\left[ r^2 + b^2 -2b r\cos(\varphi-\theta)\right].
\end{equation}
Noting that
\begin{align}
  \ln|1 - 2\epsilon \cos\alpha + \epsilon^2| 
&= \ln|1-\epsilon e^{i\alpha}| + \ln|1-\epsilon e^{-i\alpha}|\nonumber\\
&= \sum_{m=1}^\infty (-1)^{m+1}(-1)^m \frac{\epsilon^m}{m} \left(e^{im\alpha} + e^{-im\alpha}\right) \nonumber\\
&=\sum_{m=1}^\infty \frac{ -2\epsilon^m}{m} \cos(m\alpha)
\end{align}
for $\epsilon<1$, the poloidal flux can be written as a multipole
expansion
\begin{align}
 \Psi_p(r>b,\varphi) &= \frac{\mu_0I_p}{2\pi}\left[-\ln r +\sum_{m=1}^\infty \left(\frac{b}{r}\right)^m  \frac{\cos[m(\varphi-\theta)]}{m}\right]\\
 \Psi_p(r<b,\varphi) &= \frac{\mu_0I_p}{2\pi}\left[-\ln b +\sum_{m=1}^\infty \left(\frac{r}{b}\right)^m  \frac{\cos[m(\varphi-\theta)]}{m}\right]
\end{align}
The associated potential function is then shown via
(\ref{eq:cauchy-riemann}) to be
\begin{align}
  \Phi_p(r>b,\varphi) &= \frac{\mu_0 I_p}{2\pi}\left\{\varphi + \sum_{m=1}^\infty \left(\frac{b}{r}\right)^m  \frac{\sin[m(\varphi-\theta)]}{m}\right\}\label{eqn:wire_cyl_outer}\\
  \Phi_p(r<b,\varphi) &= -\frac{\mu_0 I_p}{2\pi} \sum_{m=1}^\infty \left(\frac{r}{b}\right)^m  \frac{\sin[m(\varphi-\theta)]}{m} \label{eqn:wire_cyl_inner}
\end{align}
\subsection{Solution to 2D Laplace from wire current distribution}
\label{sec:current_derivation}
The current distribution from the wire is represented as ($\hat{e}_\varphi = r \grad \varphi$)
\begin{align}
  \bm{j}_p &= I_p \delta(\bm{x}-\bm{x}_p)\bm{e}_z = \frac{I_p}{r} \delta(r-b)\delta(\varphi - \theta) \hat{\bm{e}}_r\times \hat{\bm{e}}_\varphi \nonumber\\
&= I_p\delta(r-b)\delta(\varphi - \theta)\grad r\times \grad \varphi
\end{align}
The dirac delta in angles can be represented by an infinite Fourier series as
\begin{equation}
  \delta(\varphi - \theta) \equiv \frac{1}{2\pi} + \frac{1}{\pi}\sum_{m=1}^\infty \cos[m(\varphi-\theta)]  
\end{equation}
so that
\begin{equation}
  \mu_0\bm{j}_p = \curl \left\{-\frac{\mu_0 I_p}{2\pi}\delta(r-b) \left[\varphi + 2 \sum_{m=1}^\infty \frac{\sin[m(\varphi-\theta)]}{m} \right] \grad r\right\}  
\end{equation}
Since $\mu_0 \bm{j}_p = \curl \bm{B}_p$, it is then natural to write the
magnetic field as
\begin{equation}
  \bm{B}_p = -\frac{\mu_0 I_p}{2\pi}\delta(r-b) \left[\varphi + 2 \sum_{m=1}^\infty \frac{\sin[m(\varphi-\theta)]}{m}\right]\grad r + \grad \Phi_p
 \label{eqn:bsurf}
\end{equation}
where the potential function must satisfy $\div\bm{B}_p =
\grad^2\Phi_p=0$ everywhere in the vacuum. The solution $\Phi_p^\pm =
\bar{\Phi}_p^\pm + \tilde{\Phi}_p^\pm$ is broken in the region
enclosed by the singular layer at $r=b$ (the minus solution) and
outside (the plus solution) into a component $\tilde{\Phi}_p$ that is
single-valued in $\varphi$ and a secular term $\bar{\Phi}_p\propto
\varphi$. The latter provides the current within the radius $r$
through the circulation $\oint \bm{B}_p\cdot d\bm{l}/2\pi = \int \mu_0
\bm{j}_p\cdot d\bm{\sigma}/2\pi$. Hence,
\begin{equation}
  \bar{\Phi}_p(r,\varphi) =  \frac{\mu_0 I_p}{2\pi} \varphi\Theta(r-b)
\end{equation}
where $\Theta(x)$ is the Heavyside distribution, $\Theta'=\delta$.

In the enclosed region, the single-valued solution to the 2D Laplace
equation $\tilde{\Phi}_p^-$ will have to be of the following form to
avoid singularities at $r=0$
\begin{multline}
  \tilde{\Phi}_p(r<b,\varphi) = A^- \\
+\sum_{m=1}^\infty \left(\frac{r}{b}\right)^m\left(S^-_m\sin[m(\varphi-\theta)] + C^-_m\cos[m(\varphi-\theta)]\right)
\end{multline}
and in the outer region to avoid diverging magnetic fields as
$r\rightarrow \infty$, the single-valued solution must be of the form
\begin{multline}
  \tilde{\Phi}_p(r>b,\varphi) = A^+ + B^+\ln r \\
+ \sum_{m=1}^\infty \left(\frac{b}{r}\right)^m\left(S^+_m\sin[m(\varphi-\theta)] + C^+_m\cos[m(\varphi-\theta)]\right)
\end{multline}
The constant coefficients $A^+$ and $A^-$ play no role and can be
omitted.

By virtue of $\div\bm{B}_p = 0$, the normal component of the magnetic
field $\bm{B}_p\cdot \hat{\bm{e}}_r = \partial_r \Phi_p$ is continuous across the singular
layer at $r=b$, which provides the matching conditions
\begin{align}
    B^+ &= 0 &
    S_m^+ &= -S_m^- & 
    C_m^+ &= -C_m^-
\end{align}
By integrating the continuous normal component of the magnetic field
(\ref{eqn:bsurf}) across the singular layer, another matching
condition is obtained as
\begin{multline}
  \int_{b-\epsilon}^{b+\epsilon} dr \bm{B}_p\cdot \hat{\bm{e}}_ r = -\frac{\mu_0 I_p}{2\pi}\left[ \varphi + 2 \sum_{m=1}^\infty \frac{\sin[m(\varphi-\theta)]}{m}\right]\\
 + \Phi_p(b+\epsilon,\varphi) - \Phi_p(b-\epsilon,\varphi) \overset{\epsilon\rightarrow 0}{\longrightarrow} 0
\end{multline}
which means that
\begin{align}
  C_m^+ &= C_m^-= 0  &
  S_m^+ = -S_m^- = \frac{\mu_0I_p}{2\pi m } 
\end{align}
and one obtains exactly the same solution as
(\ref{eqn:wire_cyl_outer}-\ref{eqn:wire_cyl_inner}). Notice how the
representations of the magnetic field (\ref{eqn:bnat}) and
(\ref{eqn:bsurf}) are exactly the same.
\end{document}